\documentclass[conference]{IEEEtran}
\makeatletter\if@twocolumn\PassOptionsToPackage{switch}{lineno}\else\fi\makeatother

\usepackage{graphicx}
\usepackage[T1]{fontenc}
\ifCLASSINFOpdf
\else
\fi
\usepackage{amsmath}
\usepackage[cmintegrals]{newtxmath}
\usepackage{url,multirow,morefloats,floatflt,cancel,tfrupee}
\makeatletter

\AtBeginDocument{\@ifpackageloaded{textcomp}{}{\usepackage{textcomp}}}
\makeatother
\usepackage{colortbl}
\usepackage{xcolor}
\usepackage{pifont}
\usepackage[nointegrals]{wasysym}
\makeatletter

\def\mcWidth#1{\csname TY@F#1\endcsname+\tabcolsep}

\def\cAlignHack{\rightskip\@flushglue\leftskip\@flushglue\parindent\z@\parfillskip\z@skip}
\def\rAlignHack{\rightskip\z@skip\leftskip\@flushglue \parindent\z@\parfillskip\z@skip}

\@ifundefined{etal}{}{}

\usepackage{ifxetex}
\ifxetex\else\if@twocolumn\@ifpackageloaded{stfloats}{}{\usepackage{dblfloatfix}}\fi\fi

\AtBeginDocument{
\expandafter\ifx\csname eqalign\endcsname\relax
\def\eqalign#1{\null\vcenter{\def\\{\cr}\openup\jot\m@th
  \ialign{\strut$\displaystyle{##}$\hfil&$\displaystyle{{}##}$\hfil
      \crcr#1\crcr}}\,}
\fi
}

\AtBeginDocument{%
  \@ifpackageloaded{endfloat}%
   {\renewcommand\efloat@iwrite[1]{\immediate\expandafter\protected@write\csname efloat@post#1\endcsname{}}}{\newif\ifefloat@tables}%
}%

\def\BreakURLText#1{\@tfor\brk@tempa:=#1\do{\brk@tempa\hskip0pt}}
\let\lt=<
\let\gt=>
\def\processVert{\ifmmode|\else\textbar\fi}

\@ifundefined{subparagraph}{
\def\subparagraph{\@startsection{paragraph}{5}{2\parindent}{0ex plus 0.1ex minus 0.1ex}%
{0ex}{\normalfont\small\itshape}}%
}{}

\newcommand\role[1]{\unskip}
\newcommand\aucollab[1]{\unskip}
  
\@ifundefined{tsGraphicsScaleX}{\gdef\tsGraphicsScaleX{1}}{}
\@ifundefined{tsGraphicsScaleY}{\gdef\tsGraphicsScaleY{.9}}{}
\def\checkGraphicsWidth{\ifdim\Gin@nat@width>\linewidth
	\tsGraphicsScaleX\linewidth\else\Gin@nat@width\fi}

\def\checkGraphicsHeight{\ifdim\Gin@nat@height>.9\textheight
	\tsGraphicsScaleY\textheight\else\Gin@nat@height\fi}

\def\fixFloatSize#1{}
\let\ts@includegraphics\includegraphics

\def\inlinegraphic[#1]#2{{\edef\@tempa{#1}\edef\baseline@shift{\ifx\@tempa\@empty0\else#1\fi}\edef\tempZ{\the\numexpr(\numexpr(\baseline@shift*\f@size/100))}\protect\raisebox{\tempZ pt}{\ts@includegraphics{#2}}}}

\AtBeginDocument{\def\includegraphics{\@ifnextchar[{\ts@includegraphics}{\ts@includegraphics[width=\checkGraphicsWidth,height=\checkGraphicsHeight,keepaspectratio]}}}

\DeclareMathAlphabet{\mathpzc}{OT1}{pzc}{m}{it}

\def\URL#1#2{\@ifundefined{href}{#2}{\href{#1}{#2}}}

\def\UrlOrds{\do\*\do\-\do\~\do\'\do\"\do\-}%
\g@addto@macro{\UrlBreaks}{\UrlOrds}

\edef\fntEncoding{\f@encoding}

\makeatother

\newif\ifmultipleabstract\multipleabstractfalse%
\usepackage{tabulary}
\makeatletter
\AtBeginDocument{\@ifpackageloaded{longtable}{%
\def\LT@makecaption#1#2#3{%
  \LT@mcol\LT@cols c{\hbox to\z@{\hss\parbox[t]\LTcapwidth{%
    \sbox\@tempboxa{#1{#2: } #3}%
    \ifdim\wd\@tempboxa>\hsize
      #1{#2: }\textsc{#3}%
    \else
      \hbox to\hsize{\hfil\box\@tempboxa\hfil}%
    \fi
    \endgraf\vskip\baselineskip}%
  \hss}}}
}{}}
\makeatother

   \makeatletter
  \def\fig@textbf{\textbf}
   \AtBeginDocument{\renewcommand\floatc@plain[2]{\setbox\@tempboxa\hbox{{\footnotesize#1.}\footnotesize\hskip.5em#2}%
    \ifdim\wd\@tempboxa>\hsize {\fig@textbf{\footnotesize#1.}}\footnotesize\hskip.5em#2\par
        \else\hbox to\hsize{\hfil\box\@tempboxa\hfil}\fi}}
    \makeatother
  
\usepackage{float}

\begin{document}

%


        \title{IMPROVED IMAGE CODING AUTOENCODER WITH DEEP LEARNING }
      \author{
		\IEEEauthorblockN{Licheng~Xiao}\\[-12pt]Email: david.xiao.2008@gmail.com
        \vspace*{1pc}\and 
		\IEEEauthorblockN{Hairong~Wang}
        \vspace*{1pc}\and 
		\IEEEauthorblockN{Nam~Ling}}
  


\maketitle 

\begin{abstract}
In this paper, we build autoencoder based pipelines for extreme end-to-end image compression based on Ball{\'e}'s approach\unskip~\cite{669539:15842455}, which is the state-of-the-art open source implementation in image compression using deep learning. We deepened the network by adding one more hidden layer before each strided convolutional layer with exactly the same number of down-samplings and up-samplings. Our approach outperformed Ball{\'e}'s approach, and achieved around 4.0\% reduction in bits per pixel (bpp), 0.03\% increase in multi-scale structural similarity (MS-SSIM), and only 0.47\% decrease in peak signal-to-noise ratio (PSNR),  It also outperforms all traditional image compression methods including JPEG2000 and HEIC by at least 20\% in terms of compression efficiency at similar reconstruction image quality. Regarding encoding and decoding time, our approach takes similar amount of time compared with traditional methods with the support of GPU, which means it's almost ready for industrial applications.
\end{abstract}
    


%
\IEEEpeerreviewmaketitle

\section{Introduction}
In recent years, image coding has seen lots of innovations invoked by deep learning. Some deep learning approaches have outperformed all traditional methods in terms of compression efficiency and reconstruction quality. The state-of-the art deep learning approach with open source implementation was proposed by Ball{\'e} et al in 2018\unskip~\cite{669539:15842455}, and open sourced on github in 2019\unskip~\cite{669539:15842497}. The goal of the research is to improve the approach and achieve better experimental results. \footnote{{\copyright}  2020 IEEE.  Personal use of this material is permitted. Permission from IEEE must be obtained for all other uses, in any current or future media, including reprinting/republishing this material for advertising or promotional purposes, creating new collective works, for resale or redistribution to servers or lists, or reuse of any copyrighted component of this work in other works.}

By learning from our previous research\unskip~\cite{669539:15842758}, and conducting dozens of new experiments in exploring neural networks architecture and hyper parameters space, we successfully outperformed the previous state-of-the-art image coding method (Ball{\'e} et al 2018)\unskip~\cite{669539:15842455}\footnote{The source code of our implementation is available in bit.ly/deepimagecompressiongithub.}. With similar PSNR and MS-SSIM, our approach achieved around 4\% rate saving. We also proved that replacing single convolutional layer with double convolutional layers and smaller kernels before down sampling is an effective way in improving model perforance of autoencoder models used in image compression.

For faster experiments, we implemented the functionality to encode, decode and evaluate images in batches, which is around 10 times faster than previous approaches.

Another major contribution of this paper is the accurate measurement of encoding and decoding time of the previous state-of-the-art image coding method (Ball{\'e} et al 2018)\unskip~\cite{669539:15842455}, and comparing it with that of traditional approaches like JPEG2000 and HEIC. Experimental results show that our approach only cost 12.12\% more decoding time compared with the baseline, which is quite good for 4\% rate saving. Compared with HEIC, our approach cost around 85\% more decoding time, which is not bad for 24.17\% rate saving at almost the same PSNR and MS-SSIM, and much better than most deep learning approaches so far, which often require 10 to 1000 times decoding time comparing with HEIC even with the support of GPU.
    
\section{Prior Art}
As we mentioned in our previous work\unskip~\cite{669539:15842758}, traditional image compression methods were represented by JPEG, JPEG2000\unskip~\cite{669539:15842760}, BPG\unskip~\cite{669539:16057467} and HEIC\unskip~\cite{669539:16053509}. In comparison, deep learning based methods were represented by generative adversarial networks (GAN)\unskip~\cite{669539:15842869}, super resolution\unskip~\cite{669539:15842927} and autoencoder\unskip~\cite{669539:15871638}\unskip~\cite{669539:15842455}\unskip~\cite{669539:15871679}\unskip~\cite{669539:15889776} . Among all deep learning based methods with open source implementation, the autoencoder model proposed by Ball{\'e} et al in 2018\unskip~\cite{669539:15842455} was the state-of-the-art, and outperformed all traditional and other deep learning based methods regarding reconstruction quality and compression efficiency, thus we select it as our baseline to compare.

\subsection{Baseline Overview}The baseline was built on top of Ball{\'e}'s work in 2017\unskip~\cite{669539:15871638}, and the major difference is that it incorporates a hyperprior to capture spatial dependencies in the latent representation. This hyperprior relates to side information and was trained jointly with the underlying autoencoder\unskip~\cite{669539:15842455}. This innovation yielded the state-of-the-art rate-distortion performance so far in all published ANNs (artificial neural networks) with open source implementation.

The structure of the baseline is shown in Figure~\ref{f-973a97c83eb5}. The left side shows an image autoencoder architecture. The right side corresponds to the autoencoder implementing the hyperprior. The factorized-prior model uses the identical architecture for the analysis and synthesis transforms g\ensuremath{_{a\ }}and g\ensuremath{_{s}}. Q represents quantization. AE and AD represent arithmetic encoder and arithmetic decoder, respectively. Convolution parameters are denoted as: number of filters \ensuremath{\times} kernel support height \ensuremath{\times} kernel support width / down or up sampling stride, where \ensuremath{\uparrow} indicates up sampling and \ensuremath{\downarrow} down sampling. N and M were chosen dependent on \ensuremath{\lambda }\unskip~\cite{669539:15842455}. For the baseline, we chose N = M = 192 when \ensuremath{\lambda } = 0.01.

The encoder subjects the input image x to g\ensuremath{_{a}}, yielding the responses y with spatially varying standard deviations. The responses are fed into h\ensuremath{_{a}}, summarizing the distribution of standard deviations in z. z is then quantized, compressed, and transmitted as side information. The encoder then uses the quantized vector to estimate, the spatial distribution of standard deviations, and uses it to compress and transmit the quantized image representation. The decoder first recovers from the compressed signal. It then uses h\ensuremath{_{s\ }}to obtain, which provides it with the correct probability estimates to successfully recover as well. It then feeds into g\ensuremath{_{s}} to obtain the reconstructed image.\unskip~\cite{669539:15842455}

\bgroup
\fixFloatSize{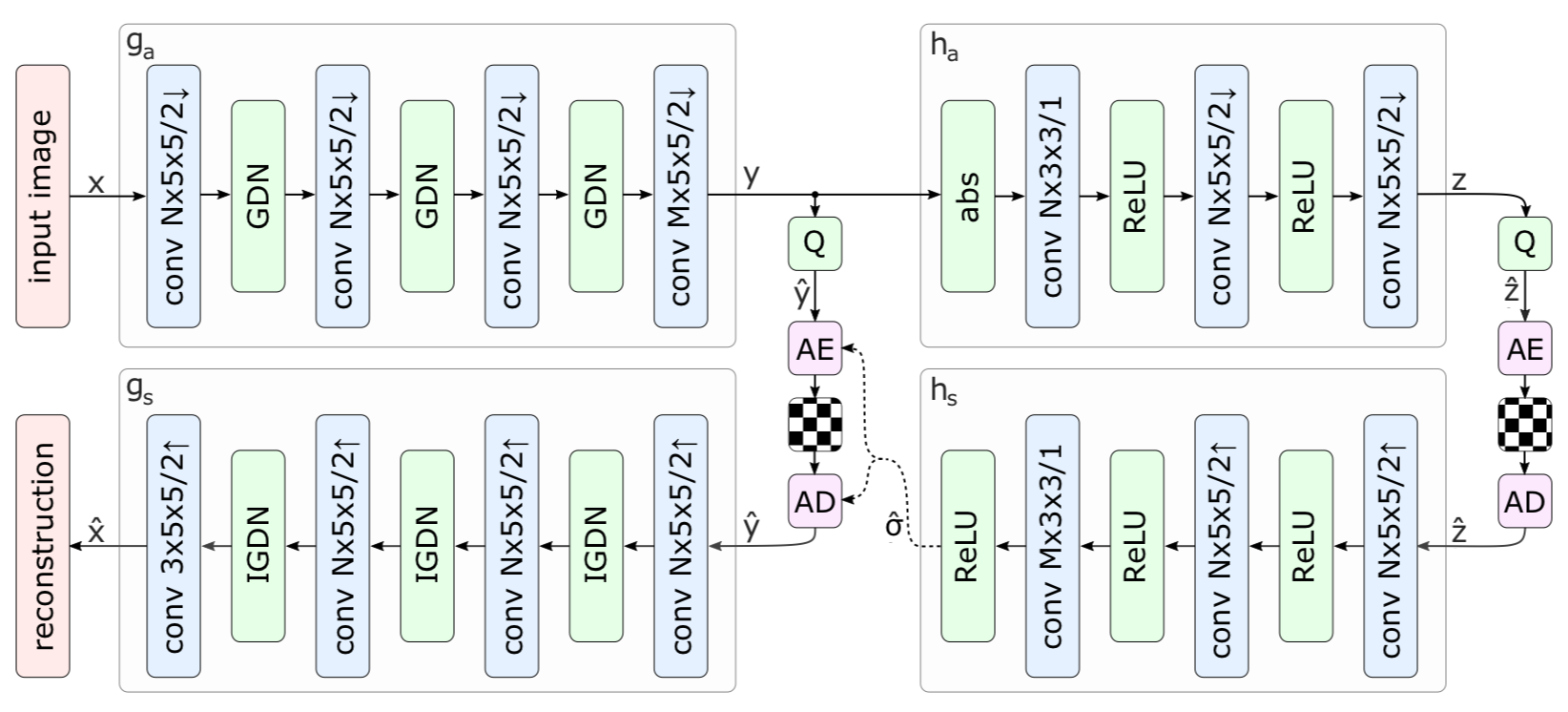}
\begin{figure}[!htbp]
\centering \makeatletter\IfFileExists{images/b769825d-2a9a-4895-8651-05db96a3b822-ustructure-of-baseline-2018.png}{\includegraphics{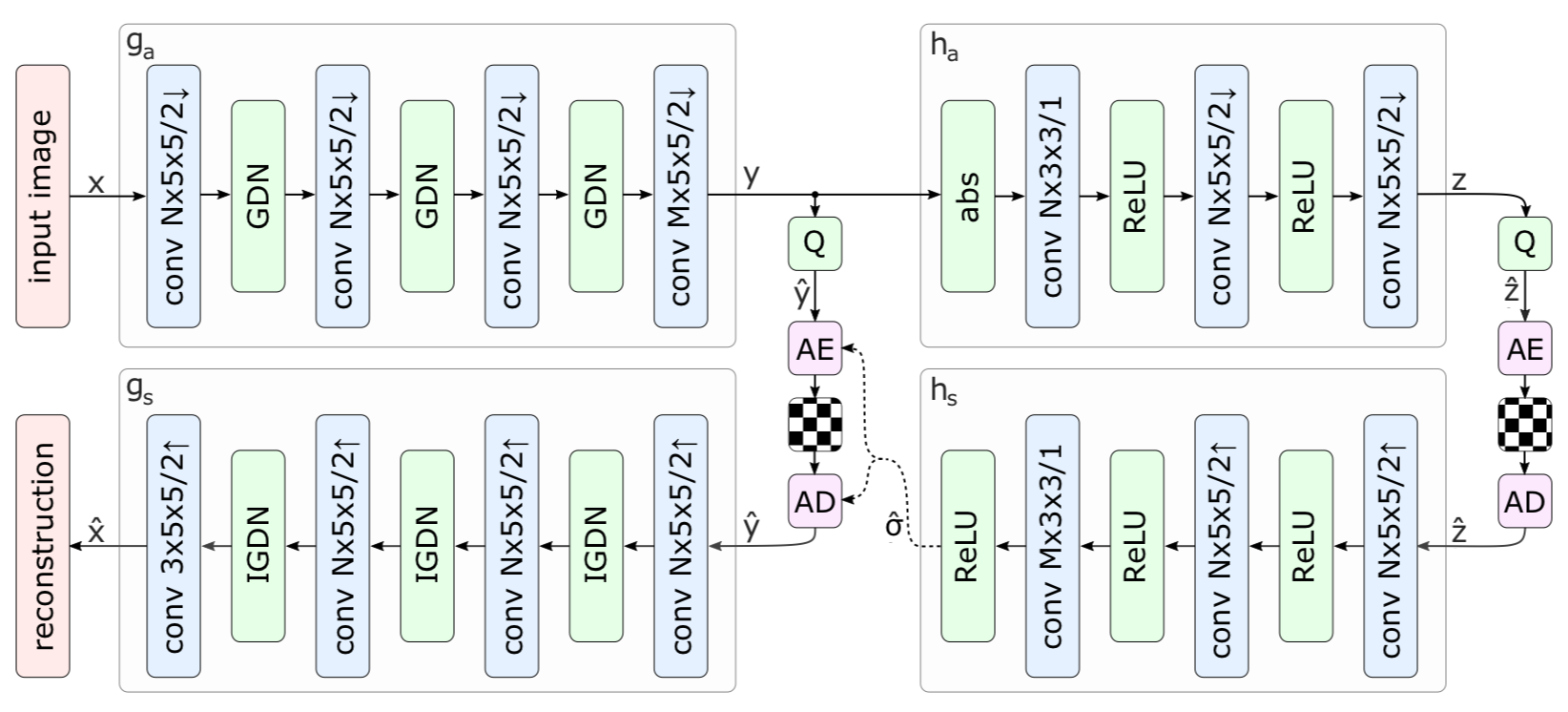}}{}
\makeatother 
\caption{{Network architecture of the baseline.\unskip~\protect\cite{669539:15842455}}}
\label{f-973a97c83eb5}
\end{figure}
\egroup

\section{Proposed Method}
During our previous work\unskip~\cite{669539:15842758}, we did many experiments, and did not record all the intermediate results, which made it difficult to isolate the influence of each variable. Therefore, when improving the baseline this time, we reduced the number of variables that we modified simultaneously, so that we can better track the differences caused by changing single variable. We used very similar network structure as the baseline, only replacing the kernel size from 5\ensuremath{\times} 5 to 3\ensuremath{\times} 3 and added one additional convolutional layer before each down sampling or up sampling. Experimental results shown that our modifications improved  compression ratio by around 4\% with similar reconstruction quality.

Similar to our modifications in previous work\unskip~\cite{669539:15842758}, we added one more convolutional layer before each down sampling or up sampling. The difference is that we used the same number of down sampling and up sampling as that of the baseline, to better isolate variables in our experiments. Since the encoder and decoder were similar in structure, we use encoder as example to illustrate our improvement as in Figure~\ref{f-37e0fde7b898}. In the encoder of Ball{\'e}'s approach, there are four convolutional layers with kernel size of 5 \ensuremath{\times} 5 and four down samplings. In our approach, there are eight convolutional layers with kernel size of 3 \ensuremath{\times} 3 and four down samplings.

\bgroup
\fixFloatSize{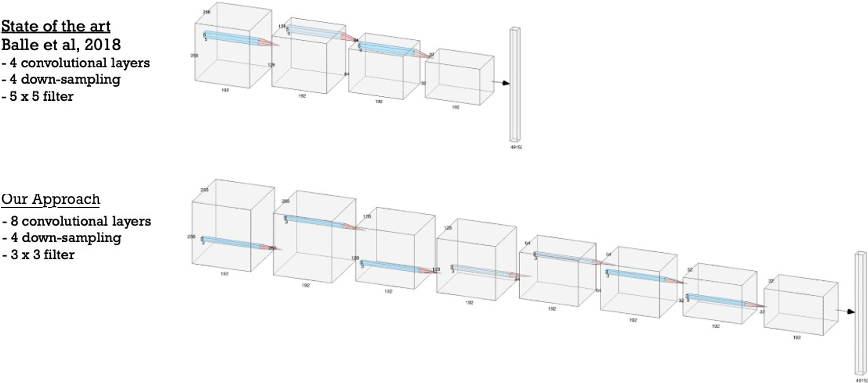}
\begin{figure*}[!htbp]
\centering \makeatletter\IfFileExists{images/2136ddd7-e404-4b26-94ed-ffafe3972f71-upicture1.png}{\includegraphics{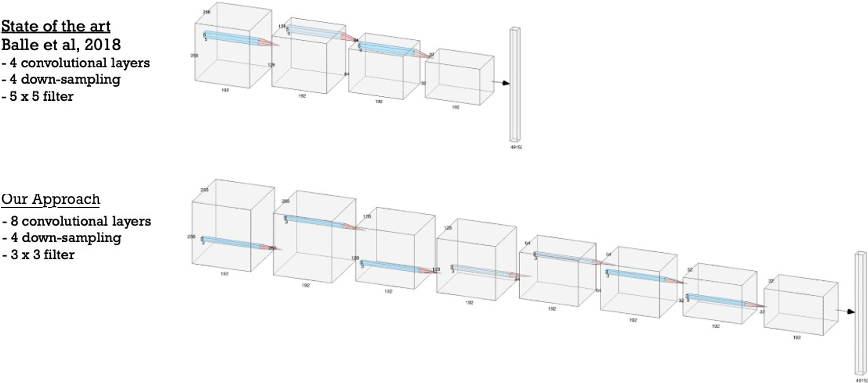}}{}
\makeatother 
\caption{{Comparison between encoder structure of Ball{\'e}'s approach and our approach.}}
\label{f-37e0fde7b898}
\end{figure*}
\egroup
This modification is the best among dozens of experiments for structural improvement. Experimental results have shown that this modification works well and can provide around 4\% improvement on bpp with similar mean square error (MSE) when the models were trained to 1 million iterations. The reason why our modifications works better is possibly that more convolutional layers before information loss caused by down sampling can provide more parameters to learn image features, so that the model can better capture details in image compression. This is similar to our improvement in previous work\unskip~\cite{669539:15842758}, but with more strict control on independent variables.

The size of kernels was reduced from 5 \ensuremath{\times} 5 to 3 \ensuremath{\times} 3 to maintain a relatively similar scale for overall convolution, since the model has two convolutional layers instead of one before each down sampling or up sampling in our approach. This should not have much influence on the result but can be isolated in future experiments.
    
\section{Experimental Results}
For the baseline, we only trained the model with lambda = 0.01 with 1M iterations. Each experiment took around 1 week on single GPU (Nvidia GeForce GTX 1070).

All models were trained using CLIC professional-train dataset\unskip~\cite{669539:15889945} and evaluated using Kodak True Color Image Suite\unskip~\cite{669539:16055445}. The CLIC professional-valid dataset contains images that are too large to fit into the eight gigabytes memory of Nvidia Geforce GTX1070, which would lead to incomplete reconstructed images and incorrect evaluation metrics. CLIC professional-train dataset contains more than 600 pictures taken by professional cameras, with resolutions from standard definition (SD) to high definition (HD), covering various scenarios. The Kodak True Color Image Suite contains pictures of different categories, including scenarios in the wild, buildings in the city, portraits, and sports, with resolution of either 512 \ensuremath{\times} 768 or 768 \ensuremath{\times} 512, in RGB color domain.

The baseline used RGB domain as the only supported color domain, and we did not change that part during all our experiments. Therefore, during preprocessing, we converted all the training images from sRGB domain to RGB domain.

Each training image was randomly sliced to patches of size 256 \ensuremath{\times} 256. Ideally, this step belongs to preprocessing and should be executed independent of the training process. However, the baseline script directly included this step in the training process, which might have some influence on duplicating experimental results. Considering the training iterations was more than 1 million, which was much larger than the number of training images, the differences caused by randomly slicing should be minimal and can be ignored. Therefore, we did not change this part in our approach during these experiments.

Among all models we trained, the best one was the one described in Section III, which added one additional convolutional layer before each down sampling or up sampling with kernel size of 3 \ensuremath{\times} 3 instead of 5 \ensuremath{\times} 5.

\subsection{Single-point comparison between the Baseline, Our approach, JPEG2000 and HEIC}We trained both the baseline and our approach to 1 million iterations and observed that our approach can achieve significantly lower bpp with similar MSE, and this trend is consistent and stable throughout the training process.

As is shown in Figure~\ref{f-216807186096}, at the beginning, the MSE of our approach is higher than that of Ball{\'e}'s approach. However, as training went on, the MSE of our approach steadily decreased to the same level as that of Ball{\'e}'s approach and reached a stable status from 600k to 1M iterations.

\bgroup
\fixFloatSize{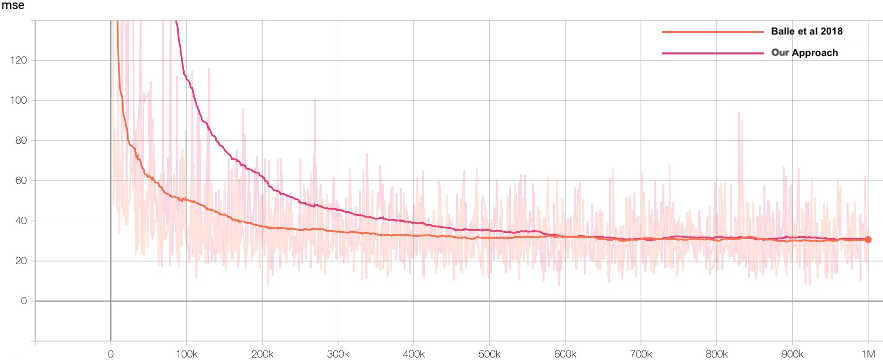}
\begin{figure}[!htbp]
\centering \makeatletter\IfFileExists{images/60817cfe-cd21-4390-9ab9-e14e5a9c4322-upicture2.png}{\includegraphics{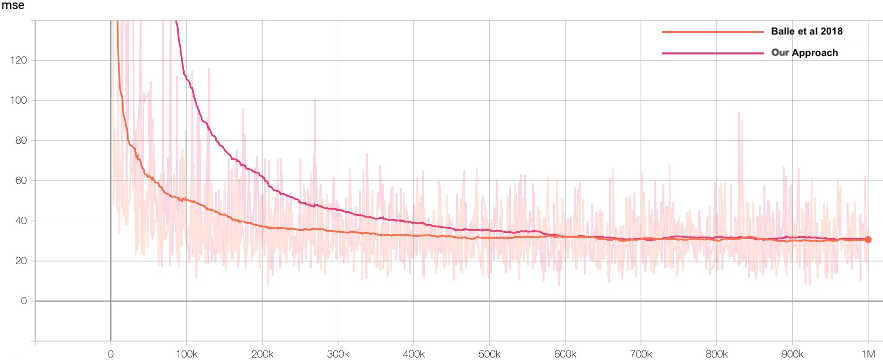}}{}
\makeatother 
\caption{{Comparison of MSE over training iterations between Ball{\'e}'s approach and our approach.}}
\label{f-216807186096}
\end{figure}
\egroup
As is shown in Figure~\ref{f-f4644ace38ec}, at the beginning, the bpp of our approach was lower than that of Ball{\'e}'s approach, and as training went on, this advantage maintained well till the end of 1 million iterations. At the end of training, our approach achieved around 4\% reduction in bpp than that of Ball{\'e}'s approach.

\bgroup
\fixFloatSize{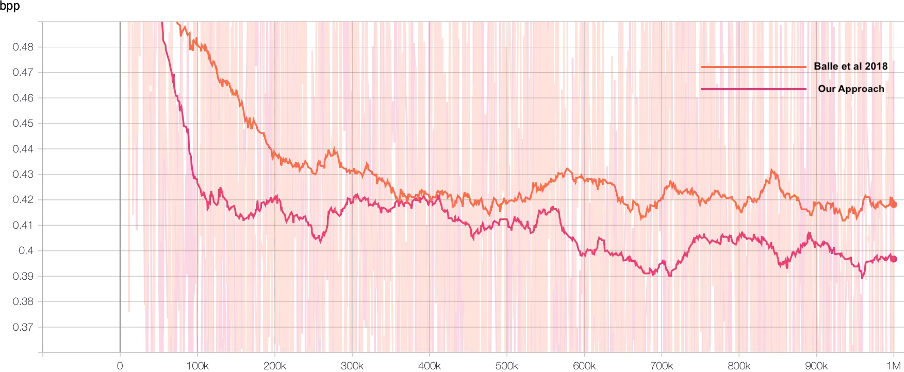}
\begin{figure}[!htbp]
\centering \makeatletter\IfFileExists{images/3.png}{\includegraphics{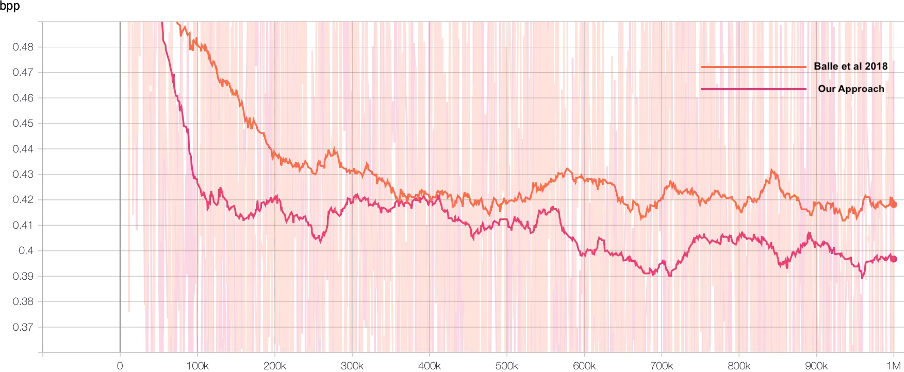}}{}
\makeatother 
\caption{{Comparison of bpp over training iterations between Ball{\'e}'s approach and our approach.}}
\label{f-f4644ace38ec}
\end{figure}
\egroup
Note that the first 50K iterations were a bit different from later iterations, as is shown in Figure~\ref{f-af7e98e3105b}. Our approach started with a lower bpp, and Ball{\'e}'s approach started with a higher bpp. The bpp of our approach then increased a bit till iteration 15k, while the bpp of the baseline decreased. From iteration 15k to 30k, the bpp of both approaches were very close to each other. After iteration 30k, the bpp of our approach started to decrease faster than that of Ball{\'e}'s approach. The reason for this difference in the starting bpp still needs further experiments to reveal.

\bgroup
\fixFloatSize{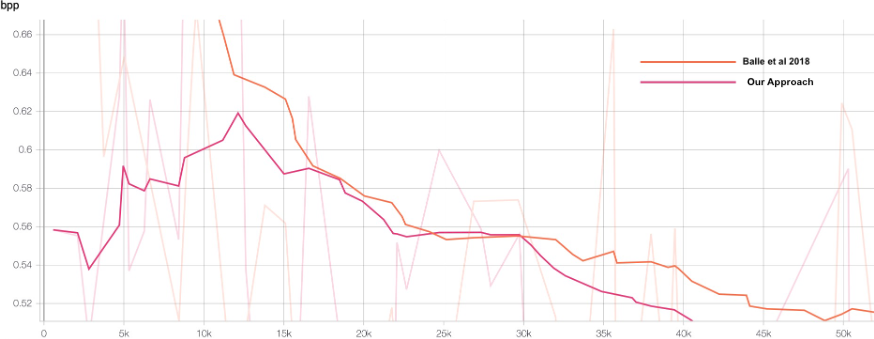}
\begin{figure}[!htbp]
\centering \makeatletter\IfFileExists{images/f1a28b7f-f34f-4d78-887e-2d3c1863d051-upicture4.png}{\includegraphics{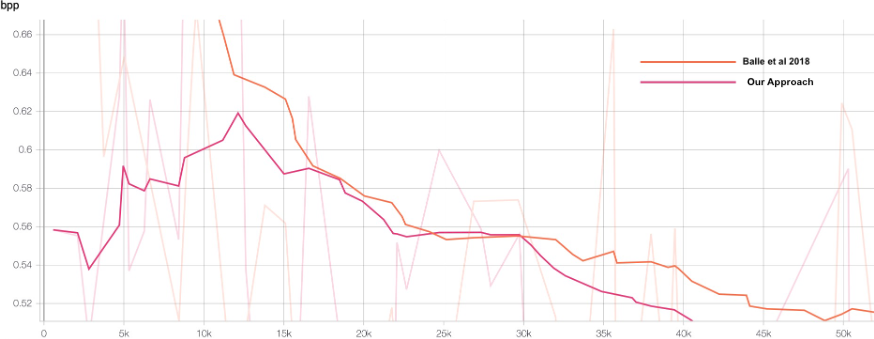}}{}
\makeatother 
\caption{{Comparison of bpp in the first 50K training iterations between Ball{\'e}'s approach and our approach.}}
\label{f-af7e98e3105b}
\end{figure}
\egroup
Averaged on all 24 pictures in Kodak True Color Image Suite, our approach achieved better performance than all previous standards, as is shown in Table~\ref{tw-02d4d74205d8}. Compared with the baseline, our approach achieved around 4.0\% reduction in bpp, 0.03\% increase in MS-SSIM and only 0.47\% decrease in PSNR. Compared with HEIC, our approach achieved around 24.17\% reduction in bpp, at almost the same PSNR and MS-SSIM.

Regarding encoding and decoding time, our approach takes 12.12\% more decoding time than the baseline, and around 85\% more decoding time compared with HEIC with the support of GPU. This is much better than most deep learning approaches so far, which often requires 10 to 1000 times more decoding time compared with HEIC.

\begin{table}[!htbp]
\caption{{Comparison between JPEG2000, HEIC, Ball{\'e}'s approach, our approach, over Kodak True Color Image Suite.} }
\label{tw-02d4d74205d8}
\def\arraystretch{1}
\ignorespaces 
\centering 
\begin{tabulary}{\linewidth}{p{\dimexpr.14\linewidth-2\tabcolsep}p{\dimexpr.14\linewidth-2\tabcolsep}p{\dimexpr.14\linewidth-2\tabcolsep}p{\dimexpr.14\linewidth-2\tabcolsep}p{\dimexpr.14\linewidth-2\tabcolsep}p{\dimexpr.14\linewidth-2\tabcolsep}p{\dimexpr.16\linewidth-2\tabcolsep}}
\hline Standard & Average \mbox{}\protect\newline bpp & Average \mbox{}\protect\newline PSNR & Average \mbox{}\protect\newline MS-SSIM & Encoding \mbox{}\protect\newline Time \mbox{}\protect\newline (seconds) & Decoding \mbox{}\protect\newline Time \mbox{}\protect\newline (seconds) & Hardware\\
\hline 
JPEG 2000 &
  0.7934 &
  32.60 &
  0.9777 &
  0.10 &
  0.12 &
  Intel Core i7 \mbox{}\protect\newline 2.8Hz \mbox{}\protect\newline 2 core\\
HEIC &
  0.5594 &
  31.89 &
  0.9693 &
  0.15 &
  0.10 &
  Intel Core i7 \mbox{}\protect\newline 2.8Hz \mbox{}\protect\newline 2 core\\
Ball{\'e} et al \mbox{}\protect\newline 2018 &
  0.4419 &
  32.03 &
  0.9674 &
  0.1222 &
  0.1650 &
  single Nvidia \mbox{}\protect\newline Geforce GTX \mbox{}\protect\newline 1070\\
Our Approach &
  0.4242 &
  31.88 &
  0.9677 &
  0.2369 &
  0.1850 &
  single Nvidia \mbox{}\protect\newline Geforce GTX \mbox{}\protect\newline 1070\\
\hline 
\end{tabulary}\par 
\end{table}
Note that this is the minimum bpp that JPEG2000 could achieve with the evaluation images on macOS Mojave 10.14.6.

Besides, the encoding time and decoding time for JPEG2000 and HEIC were estimated manually by encoding and decoding all 24 images in Kodak True Color Image Suite on a MacBook Pro laptop. The actual time should be slightly shorter since the I/O time of SSD (solid state drive) was not subtracted.

For the baseline, the official script did not include timing options, but we added it by ourselves to the baseline and our approach. We pre-loaded the model before encoding and decoding images and excluded I/O time when calculating encoding and decoding time for the baseline and our approach.

Note that for some large pictures in CLIC professional-valid dataset, the encoding time for our approach would be significantly prolonged due to exceeding GPU memory limit. 

When compressing some large images, the baseline and our approach might raise alarm when GPU memory is not sufficient. This usually would not cause problem if each image was compressed using separate Python session but might cause problem if many images were compressed one after another in the same Python session. Typical problem was that the reconstructed images were incomplete and had meaningless blocks as in Figure~\ref{f-9e59ce728450} . This phenomenon might also happen when there is no alarms at all, even ocassionally on small images. It might be caused by random computation error of GPU.

\bgroup
\fixFloatSize{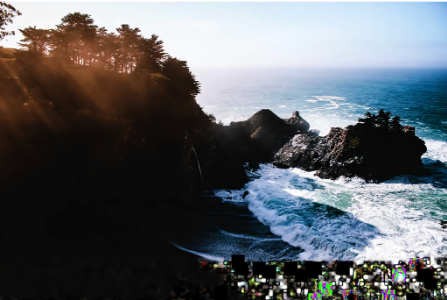}
\begin{figure}[!htbp]
\centering \makeatletter\IfFileExists{images/1ec570c5-3f04-4da8-841a-35f7e4344fb8-upicture5.png}{\includegraphics{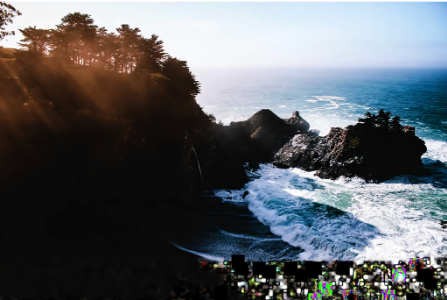}}{}
\makeatother 
\caption{{An incomplete picture processed by Ball{\'e}'s approach. Right-bottom corner was incomplete due to exceeding GPU memory limit, competing with other tasks in the same Python session or random computation error of GPU.}}
\label{f-9e59ce728450}
\end{figure}
\egroup

\subsection{Other Observations}Pictures with alpha channel can't be directly compressed using the baseline or our approach. The alpha channel might be missing in the reconstructed images, which would make the reconstructed images look significanly darker than the original ones.

Another interesting observation is that the image compressed by HEIC sometimes might lose one pixel in height or width when compressing images with dimension of odd numbers, resulting in a resolution different from that of the original image. This change in resolution was unexpected, but we got the same result after repeating the operation several times on macOS Mojave 10.14.6.
    
\section{Conclusion and Future Scope}
In our improvement on Ball{\'e}'s approach\unskip~\cite{669539:15842455}, which is the state-of-the-art image compression approach using deep learning with open source implementation, we proposed an improved autoencoder that outperformed JPEG2000, HEIC, and Ball{\'e}'s approach in bpp at comparable PSNR and MS-SSIM. 

Compared with Ball{\'e}'s approach, our approach achieved around 4.0\% reduction in bpp, 0.03\% increase in MS-SSIM and only 0.47\% decrease in PSNR. 

Compared with HEIC, our approach achieved around 24.17\% reduction in bpp, at almost the same PSNR and MS-SSIM. 

Regarding encoding and decoding time, our approach takes 12.12\% more decoding time than the baseline, and around 85\% more decoding time compared with HEIC with the support of GPU. This is much better than most deep learning approaches so far, which often requires 10 to 1000 times more decoding time compared with HEIC.

In conclusion, we successfully achieved the research goal by improving the state-of-the-art image compression method and achieving the new state-of-the-art results.

We truly believe that deep learning based approaches still have lots of potentials in image and video compression, and we plan to dedicate more time in improving our approach further in the future.



%

\bibliographystyle{IEEEtran}

\bibliography{\jobname}
\vfill
\end{document}